\documentclass{PoS}
\usepackage[authoryear,square]{natbib}
\bibpunct{(}{)}{;}{a}{}{,}

\title{The Interplay between SF and AGN Activity, and its role in Galaxy Evolution.}

\ShortTitle{The SF/AGN Inteplay}

\author{
\speaker{Kim McAlpine}$^1$, Isabella Prandoni$^{2}$, Matt Jarvis$^{3,1}$, Nick Seymour$^{4}$, Paolo Padovani$^{5}$, Philip Best$^{6}$,
Chris Simpson$^{7}$, Daria Guidetti$^{2}$, Eric Murphy$^{8}$, Minh Huynh$^{9}$, Mattia Vaccari$^{1}$, Sarah White$^3$, Rob Beswick$^{10}$, Jose Afonso$^{11,12}$, Manuela Magliocchetti$^{13}$ and Marco Bondi$^{2}$
\\
$^1$University of the Western Cape,Robert Sobukwe Road, Bellville, 7535, South Africa; 
$^{2}$INAF - Istituto di Radioastronomia, via Gobetti 101, 40129 Bologna, Italy; 
$^{3}$Oxford University, Astrophysics, Department of Physics, Keble Road, Oxford OX1 3RH; 
$^{4}$CSIRO Astronomy and Space Science, P.O. Box 76, Epping, NSW 1710, Australia;
$^{5}$ESO,Karl-Schwarzschild-Str. 2, D-85748 Garching bei M\"{u}nchen, Germany; 
$^{6}$Institute for Astronomy, University of Edinburgh, Royal Observatory Edinburgh, Blackford Hill, Edinburgh EH9 3HJ; 
$^{7}$Astrophysics Research Institute, Liverpool John Moores University, Twelve Quays House, Egerton Wharf, Birkenhead CH41 1LD;
$^{8}$California Institute of Technology, IPAC, MC 220-6, Pasadena, CA 91125, USA;
$^{9}$ICRAR - University of Western Australia, M468, 35 Stirling Hwy, Crawley WA 6009, Australia;
$^{10}$University of Manchester,Macclesfield, Cheshire, SK11 9DL;
$^{11}$ Instituto de Astrof\'{i}sica e Ci\^{e}ncias do Espa\c co, Universidade de Lisboa, OAL, Tapada da Ajuda, PT1349-018 Lisboa, Portugal;
$^{12}$ Departamento de F\'{i}sica, Faculdade de Ci\^{e}ncias, Universidade de Lisboa, Edif\'{i}cio C8, Campo Grande, PT1749-016 Lisbon, Portugal;
$^{13}$ INAF-IAPS,Via Fosso del Cavaliere 100, I-00133 Roma, Italy; 
\\
E-mail: \email{kim.mcalpine at gmail.com}
}

\abstract{It has become apparent that active galactic nuclei (AGN) may have a significant impact on the growth and evolution of their host galaxies and vice versa but a detailed understanding of the interplay between these processes remains elusive. Deep radio surveys provide a powerful, obscuration-independent tool for measuring both star formation and AGN activity in high-redshift galaxies. Multiwavelength studies of deep radio fields show a composite population of star-forming galaxies and AGN, with the former dominating at the lowest flux densities (S$_{1.4\mathrm{GHz}}<$100~$\mu$Jy). The sensitivity and resolution of the SKA will allow us to identify, and separately trace, the total star formation in the bulges of individual high-redshift galaxies, the related nuclear activity and any star formation occurring on larger scales within a disc. We will therefore gain a detailed picture of the apparently simultaneous development of stellar populations and black holes in the redshift range where both star-formation and AGN activity peak (1$\leq$z$\leq$4). In this chapter we discuss the role of the SKA in studying the connection between AGN activity and galaxy evolution, and the most critical technical requirements for such of studies.}

\FullConference{
Advancing Astrophysics with the Square Kilometre Array\\
June 8-13, 2014\\
Giardini Naxos, Sicily, Italy}

\newcommand{\skipthis}[1]{}
\newcommand{\nar}{New Astronomy Reviews}
\newcommand{\apj}{ApJ}
\newcommand{\apjl}{ApJ}
\newcommand{\aj}{AJ}
\newcommand{\mnras}{MNRAS}
\newcommand{\apjs}{ApJS}
\newcommand{\aap}{A\&A}
\newcommand{\araa}{ARA\&A}
\newcommand{\nat}{Nature}
\newcommand{\arcsec}{\mbox{$^{\prime \prime}$}}

\begin{document}
\section{Introduction}

A wide range of theoretical and observational considerations point to a close connection between star-formation and AGN activity and yet a clear understanding of this complex relationship and its true role in shaping galaxy evolution remains elusive. From a theoretical perspective galaxy formation models require a mechanism to prevent the formation of cooling flows and thereby inhibit star-formation in massive early-type galaxies,  otherwise gas is channelled to the centre of the galaxy triggering star formation. Consequently the models predict overly massive and actively star-forming galaxies at low redshifts, in contrast to observations \citep[e.g.][]{White1991}. A related, similar problem is encountered on a larger scale in observations of the intracluster medium (ICM) of galaxy clusters where the X-ray emitting gas at the centre of clusters is observed to be much hotter than expected given the radiative cooling times of the system \citep[see][for a review]{Voit2005}.  A variety of semi-analytic \citep{Granato,Bower,Croton,Somerville} and hydrodynamic simulations \citep{Gabor,Dubois,Puch} have thus included `feedback' from the central AGN as a means to disrupt the predicted cooling flows and are subsequently able to better reproduce the bright end of the galaxy mass function as well as the emergence of the red sequence. Observationally the existence of correlations between the mass of the central black hole and properties of the host galaxy such as the stellar mass of the central bulge (M$_\mathrm{BH}$-M$_\mathrm{bulge}$; \citealp{Mag,Har,Scott}) and its velocity dispersion (M$_\mathrm{BH}$-$\sigma$; \citealp{Geb,Tremaine}) seem to confirm the possibility of a link between AGN and star-formation activity.

\subsection{`Quasar' Feedback}
There are two proposed modes of AGN `feedback', the first, often referred to as `quasar' mode feedback, is associated with classical luminous quasars as well as less powerful optical and X-ray bright AGN.  In this mode the AGN is powered by radiatively efficient accretion of cold gas via an accretion disk and emits powerfully across a wide range of the electromagnetic spectrum (UV through to X-ray). A dusty torus structure surrounding the black hole and accretion disk obscures the emission at some wavelengths at large polar angles \citep[][and references therein]{Antonucci}. The majority of optically selected AGN are faint at radio wavelengths and are referred to as radio quiet (RQ) AGN, but a small fraction are radio-loud and emit large scale, relativistic radio jets. The radio-loud fraction depends on mass  \citep{Best2005,Best2007}, accretion type \citep{Janssen} and redshift \citep{Donoso}. In the local universe ($0.03<$z$<0.3$) `quasar' mode accretors increase their radio-loud fraction as a shallow function of host galaxy mass (f$^{\rm{QSO}}_{\rm{RL}}\propto$M$_{*}^{1.5}$), from approximately $\sim$0.002\% at 10$^{10.7}$M$_{*}$ to 0.1\% at 10$^{11.7}$M$_{*}$ \citep{Janssen}.

Feedback occurs as a result of high-velocity winds accelerated by the AGN, either as a result of thermal heating of gas or radiation pressure on dust, which remove gas from the galaxy. Star formation will cease abruptly once the available fuel supply is removed. It is typically assumed that the amount of energy supplied to this wind `feedback' is proportional to the mass and luminosity of the central black hole, thus feedback will only become effective above a certain mass threshold, where the momentum of the outflow is large enough to overcome the gravitational potential of the black hole and halt accretion. Observationally there is little direct evidence that typical AGN, (e.g. Seyferts) in the local universe produce the required large-scale outflows. However, more promisingly, high-velocity winds have been detected in a small number of very powerful AGN, $>$10$^{45}$~erg~s$^{-1}$, in both ionised \citep{Maiolino,Pounds,Liu} and molecular gas \citep{Sturm,Cano-Diaz,Cicone2014}. These have typical speeds of 400--1000~km~s$^{-1}$ which, with some assumptions, translates to outflow rates of 700-1000's~M$_\odot$~yr$^{-1}$, which may be capable of producing the required feedback effect. The inferred outflow rates in these more powerful AGN systems are very impressive but some recent simulations suggest that they may still have little overall effect on the gas in the galactic disk. This is due to the highly variable nature of AGN activity, such that the time-averaged outflow rates may be significantly lower than instantaneously observed outflow rates \citep{Gabor}.  More observations are thus required to understand the prevalence and/or longevity of these outflows, as they have only been detected in a small number of objects thus far, as well as to establish how their influence varies as a function of accretion rate, black hole mass, stellar mass and cosmic epoch.

\subsection{`Radio' Feedback}
A second model of feedback suggests that energy injected from AGN in the form of jets is responsible for switching off cooling at the centre of massive halos. A fairly large fraction, $\sim$ 10\%,  of massive galaxies (particularly galaxies near the centers of groups and clusters) are seen to emit radio synchrotron emission from lobes powered by jets \citep{Best2005,Best2007} . Most of these radio sources have characteristics which suggest they are experiencing a mode of accretion which is distinct from that powering traditional optical quasars. Specifically they do not have emission lines characteristic of classical optical or X-ray bright quasars \citep{Best2005,Kauffmann2008}, there is no evidence of accretion related X-ray or optical emission, nor any mid-infrared emission associated with a dusty, obscuring torus \citep{Hardcastle2007}. They are radiatively inefficient with accretion rates that are believed to be much lower than in typical quasar mode accretion, i.e. they typically accrete at $<$1\% of Eddington compared to 1--10\% of Eddington for efficient accretors. These AGN are often labelled `radio' mode or `jet' mode accretors. The fraction of `radio' mode accretors increases very sharply with host galaxy mass (f$^{\rm{Radio}}_{\rm{RL}}\propto$M$_{*}^{2.5}$) and appears to saturate at $\sim$10\% at M$_*^{11.5}$ \citep{Best2005,Janssen}. The total energy output from these AGN is lower than in the radiatively efficient case. However, depending on how efficiently the kinetic energy in the jets is converted to heat in the interstellar gas, these AGN may still have considerable potential to influence the star-formation properties of the galaxies they inhabit. Observationally the strongest direct evidence in support of `radio' mode AGN feedback is the presence of bubbles and cavities in the diffuse X-ray halos of clusters, groups and individual elliptical galaxies \citep[e.g.][]{Pedlar,McNamara,Fabian, Birzan,Rafferty,Croston,Cav} which appear to be aligned with the axis of an AGN radio jet. Further evidence of the interaction between radio galaxies and their immediate environments is provided by radio polarization studies of such jet-cavity systems. These studies  demonstrate that this interaction affects not only the density distribution of the ICM/IGM but also the magnetisation of the surrounding plasma, draping the magnetic field lines around the leading edge of the radio lobes. There is thus a clear response in the external medium linked to the direction of the radio source expansion \citep{Guidetti2011,Guidetti2012}.

Estimates of the amount of energy required to inflate these bubbles can be used to infer the mechanical heating power of radio jets.  \citet{Best2006} demonstrated that the inferred  time-averaged energetic output of `radio' mode accretors, from such cluster observations, should indeed be sufficient to counter cooling losses in massive, red galaxies. \citet{Simpson2013} find evidence to suggest that this balance is still in effect at z$>$~1, implying that `radio' mode feedback is already suppressing star-formation at moderate redshifts.   

\subsection{`Quasar' versus `Radio' mode}
The reason for the fundamental differences between the two classes of AGN activity is still uncertain. It has been argued that it may arise from differences in the fuelling gas, where the radiatively inefficient `radio' mode accretors are fuelled by hot gas found in the halos surrounding the galaxy or cluster either directly via Bondi accretion \citep{Allen2006,Hardcastle2007}, or via indirect chaotic accretion of molecular gas clumps which cool out of the hot gas phase \citep{Pizzolato2010,McNamara2011,Gaspari2013}. In contrast `quasar' mode accretors are fuelled predominantly by cold gas, supplied either by mergers or by non-axisymmetric structures within the galaxy which drive disk gas to the galaxy centre \citep{Kormendy2004,Kim2012}.  Alternatively, the two modes may largely arise due to the differences in their Eddington-scaled accretion rate onto the black hole.

There are still many unanswered questions regarding the role of both `radio' and `quasar' mode accretors in galaxy evolution and their relative importance as a function of both epoch and environment. Current simulations suggest that 'quasar' mode accretion  may be the dominant feedback mechanism at high redshifts ($z>$1) where the supply of cold gas for fuel is greater. `Radio' mode feedback appears to be relatively more important at late times, where it is essential to suppress star-formation in massive elliptical galaxies in order to reproduce their observed colours in the local universe \citep[e.g.][]{Somerville}. 
\begin{figure}[tbp]
\centering 
\resizebox{8cm}{!}{\includegraphics{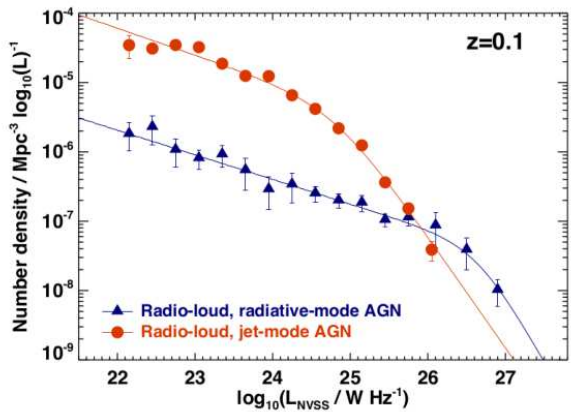}}
\hfill
\parbox[b]{55mm}{
\caption{The local radio luminosity function (0.01$<$z$<$0.3) at 1.4~GHz derived separately for radio-loud `quasar' (radiative) and `radio' (jet) mode accretors. Figure from \citet{Heckman2014}. }
\label{fig:llum}}
\end{figure}

\section{The AGN, star-formation, merger connection}

 Many theoretical models suggest that radiatively efficient `quasar' mode AGN are triggered by mergers which can drive rapid inflows of gas which fuel both intense star-formation and rapid black hole accretion \citep[e.g.][]{Barnes1992,Barnes1996,Springel,Dimatteo}. However the majority of moderate luminosity AGN, with X-ray luminosities L$_X<10^{44}$~erg~s$^{-1}$, show no evidence of merger activity  and are associated with disk host morphologies \citep{Cis,Sch,Koc}. Several studies have concluded that AGN activity appears to increase in merging systems only as a result of underlying correlations between mergers and star formation activity and between star-formation and AGN activity. In other words AGN activity remains fixed at a given specific star-formation rate while  star-formation activity generally increases as a result of mergers \citep{Reichard2009,Lib,Lia}. These results suggest that secular processes may be much more important in driving black hole growth than previously assumed, even at high redshifts (z$\sim$2) and  that the only requirement for AGN triggering is an abundant supply of central cold gas, regardless of its origin. Recent simulations have successfully produced stochastic, secularly evolving AGN where accretion is triggered by random collisions of interstellar gas clouds with the central black hole, the accretion rate and lifespan of the AGN in this scenario is strongly influenced by the gas fraction of the galaxy \citep{Gabor2013}.
 
 Major mergers do appear to be important for triggering the highest luminosity AGN with the highest accretion rates. The majority of AGN with X-ray luminosities L$_{\mathrm{X}}>10^{44}$~erg~s$^{-1}$ are hosted in galaxies with evidence of disturbed morphologies \citep{Urrutia2008,Treister2012}. Furthermore post-starbust systems, whose starbursts are possibly induced by mergers, appear to host the highest luminosity AGN \citep{Wild2007}. As the incidence of mergers and detections of potential AGN 'quasar' feedback winds are both higher in more powerful AGN it has been speculated that merger triggered AGN may be associated with stronger AGN feedback effects than AGN accreting via secular processes \cite[e.g.][]{Heckman2014}. The high sensitivity of the SKA to both obscured AGN and star-formation activity will allow us to investigate where, in terms of accretion and star-formation rate, this change in the dominant mode of black hole growth occurs and whether it is associated with a change in the interaction between star-formation and AGN activity.
 
\section{Jet induced star-formation}

The majority of theoretical models require that AGN activity results in an overall suppression of star-formation activity however there is observational evidence suggesting that AGN jets may in fact induce or enhance star-formation in some objects (i.e. positive `feedback'). This jet induced star-formation has been observed directly in only a few objects, locally in Minkowski's object \citep{Croft2006} and Centaurus A \citep{Mould2000}, at intermediate redshifts in PKS2250-41 \citep{Inskip2008} and high redshifts in  4C 41.17 \citep{Dey1997,Bicknell2000,Steinbring2014} and can take place in both FRI and FRII objects. There is also some indirect evidence of young stellar populations and  cold molecular gas being associated, and in some cases aligned, with radio jets \citep{Klamer2004,Emonts2014}.  

Simulations suggest that shocks generated by the jet propagating through a clumpy, homogeneous medium can trigger the collapse of overdense clouds resulting in actively star-forming regions \citep{Fragile2004,Gaibler2012}. It has been suggested that particularly in the early Universe, when galaxies were still forming and gas densities were much higher, jet-induced star formation may have been more common. The resolution and sensitivity of the SKA, in combination with other instruments such as ALMA, will allow detailed studies of the interaction of radio jets with the clumpy ISM and IGM of galaxies and clusters to gain clearer insights into this feedback mechanism and its relevance to galaxy evolution. 

\section{Jet induced outflows}

An interesting object, which presents an alternative model of AGN feedback, is the ULtraluminous Infrared Galaxy (ULIRG) 4C12.50, which hosts a young AGN. In this object a very fast outflow, 1000~km~s$^{-1}$ has been detected in HI, as well as in ionized \citep{Holt} and molecular gas \citep{Dasyra2012}. High resolution VLBI observations of HI demonstate that the outflow is, in projection, cospatial with a bright, highly polarized, hotspot in radio continuum located at the edge of the radio AGN jet \citep{Morganti2013}. This argues that the outflow is driven by the interaction between the radio jet and a dense cloud in the ISM rather than by an AGN wind. The energy of this outflow is less than required by most `quasar' feedback models, but may be consistent with a 'two-stage' model of feedback whereby an initially weak wind expands and dilutes dense gas clouds in the ISM making them more susceptible to secondary radiative feedback from the quasar \citep{Hopkins2010}. 

\section{The role of the SKA}

Radio surveys in the SKA era are ideal tools to investigate the SF-AGN connection for four main reasons: i) radio surveys are sensitive to emission from star-forming galaxies \citep[see e.g.][]{condon1992}, and the contribution from these star-forming galaxies increases significantly at $\mu$Jy flux densities; ii) they can detect radio emission from both `quasar' and `radio' mode accretors,  \citep[see e.g.][]{Best2012}; iii) they can provide sub-arcsec resolution essential for disentangling emission from star-formation and AGN activity; iv) radio waves offer the advantage of being unaffected by dust extinction and obscuration by circumnuclear gas, thus they are able to detect star-formation and AGN activity which would be largely obscured at optical, near-infrared and X-ray wavelengths.  The true prevalence of obscured AGN activity at moderate to high redshifts is still largely unclear. X-ray stacking of sources with excess mid-infrared {\em Spitzer} emission has revealed a population of largely unknown, heavily obscured AGN at z$\sim$2, indicating that obscured AGN activity may be much more common at high redshift than previously thought \citep{Daddi2007}. There is also evidence of a large contribution from obscured AGN activity in recent multiwavelength studies of {\em Herschel} sources \citep{Delvecchio}. The SKA will thus be extremely valuable in revealing the true incidence and relevance of such obscured AGN activity for `feedback' models at a crucial epoch of massive galaxy formation.

At frequencies of $\sim$1~GHz radio emission can be used to provide a reasonably accurate measure of galaxy star-formation rate (SFR) via the Far-Infrared Radio Correlation (FIRC; \citealp{Yun}). The relation holds over nearly 5 orders of magnitude in star-formation rate, although both observations (\citealp{Bell2003}; Jarvis et al. in prep) and theory \citep{Lackia} suggest that the relationship between star-formation rate and both radio and far-infrared emission becomes non-linear at low stellar masses and/or low star-formation rates (see also Beswick et al. this volume).  It has been confirmed that the correlation holds out to redshifts of z$\sim$1 \citep{Garrett2002,Appleton,Beswick2008,Mao2011,Murphyb}, and even as high as z$\sim$2 \citep{Sargent,Bourne,Thomson} there is little evidence of strong evolution in the relation. However, towards higher redshifts theory predicts that the non-thermal emission from radio galaxies will be suppressed due to inverse Compton (IC) losses off the cosmic microwave background \citep[see e.g.][]{Murphya,Lackib}, reducing the reliability of the FIRC as a star-formation estimator. 

The radio-loud AGN population, which dominates at flux densities greater than a few mJy, is comprised of both `quasar' and `radio' mode accretors.  Radio jets are associated with a small fraction of optically selected, radiatively efficient AGN and radio emission is the most reliable way of selecting samples of `radio' mode, radiatively inefficient AGN as their emission at UV and X-ray wavelengths is much weaker than in the radiatively efficient case. The `quasar' and `radio' mode AGN  dominate the radio-loud AGN population at high and low luminosities respectively. However there appear to be examples of both types of objects at all radio luminosities; the local luminosity function for radio-loud `quasar' and `radio' mode accretors are presented in figure ~\ref{fig:llum}.  Separating the contributions of the `radio' and `quasar' accretors will  thus require data at other wavelengths as neither jet morphology nor luminosity are reliable discriminants in the radio.

At $\mu$Jy levels, there is an increasing contribution to the radio population from radio-quiet (RQ) AGN \citep[e.g.][]{Simpson2006,Smolcic2008,Padovani2009,Bonzini2012}. RQ AGN are 'quasar' mode accretors which don't display the large, kiloparsec scale radio jets detected in radio loud AGN.  Hence they show the presence of AGN activity in one or more bands of the electromagnetic spectrum (e.g. optical, mid-infrared, X-ray), but their radio emission is much fainter, relative to the optical than in the traditional radio-loud case. There is intense debate regarding the dominant mechanism generating radio emission in these objects, as discussed in detail in Orienti et al. (this volume) and Smol{\v c}i{\'c} et al. (this volume).  The radio emission from most RQ AGNs is unresolved or barely resolved at few arcsec scales, indicating that the radio emission is confined in small regions (at most comparable to the host galaxy size). RQ AGNs may be scaled down versions of radio loud AGNs displaying mini radio jets, either associated to systems with very low accretion rates \citep[e.g.][]{Giroletti2009,Prandoni2010} or to efficiently accreting quasar-like systems \citep[see modelling work by][]{Jarvis2004}. In the local universe AGN core and jet structures have been observed at very high resolutions in a small number of RQ AGNs, while others find evidence to suggest that the radio emission is generated primarily by star formation in the host galaxy \citep[e.g.][]{Kimball2011,Padovani2011,Bonzini2013}, particularly at high redshifts. In the local universe both radio-quiet and radio-loud `quasar' mode accretors are associated with ongoing star-formation, making it difficult disentangle the contribution from star-formation and AGN activity in these objects. The SKA will enable us to establish whether/if radio-quiet AGN influence the gas of their host galaxy primarily via the 'quasar' wind mode of feedback, or whether small-scale radio jets in these systems also contribute some measure of 'radio' mode feedback.

\begin{figure}[tbp]
\centering 
\resizebox{8cm}{!}{\includegraphics{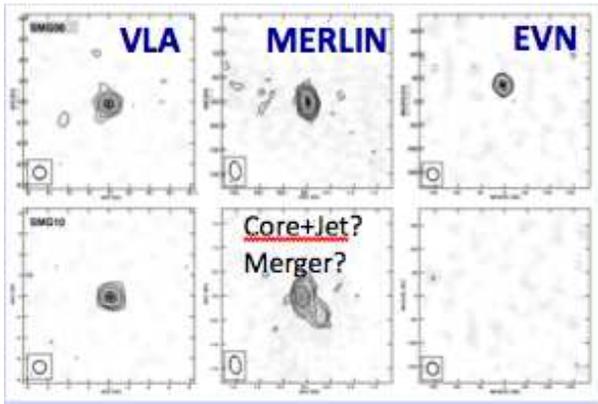}}
\hfill
\parbox[b]{55mm}{
\caption{Radio emission at 1.4 GHz on arcsec scale (VLA, left), on 200 mas scale  (MERLIN, middle) and mas scale (EVN, right) for two galaxies in the HDF North field, at $z\sim 2.7$ (top panels) and at $z\sim 1.2$ (bottom panels) respectively. An AGN core is detected in the upper object at VLBI scale, while no VLBI detection is found for the other, supporting a star-formation scenario, over a jet and core one. Figure from \cite{Biggs2010}. }
\label{fig:merlin}}
\end{figure}

\begin{figure}[tbp]
\centering 
\resizebox{8cm}{!}{\includegraphics[width=0.55\textwidth]{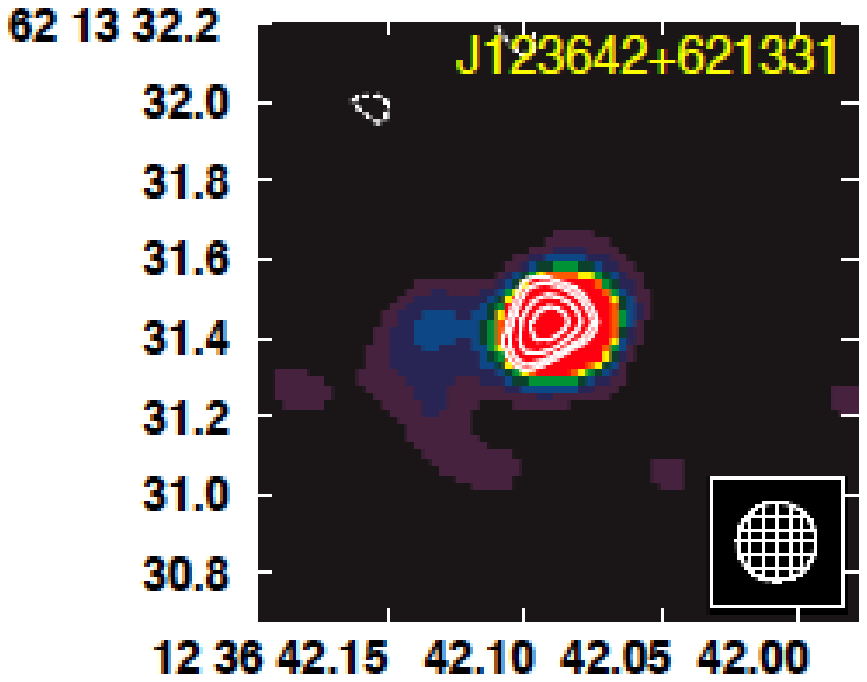}\includegraphics[width=0.45\textwidth]{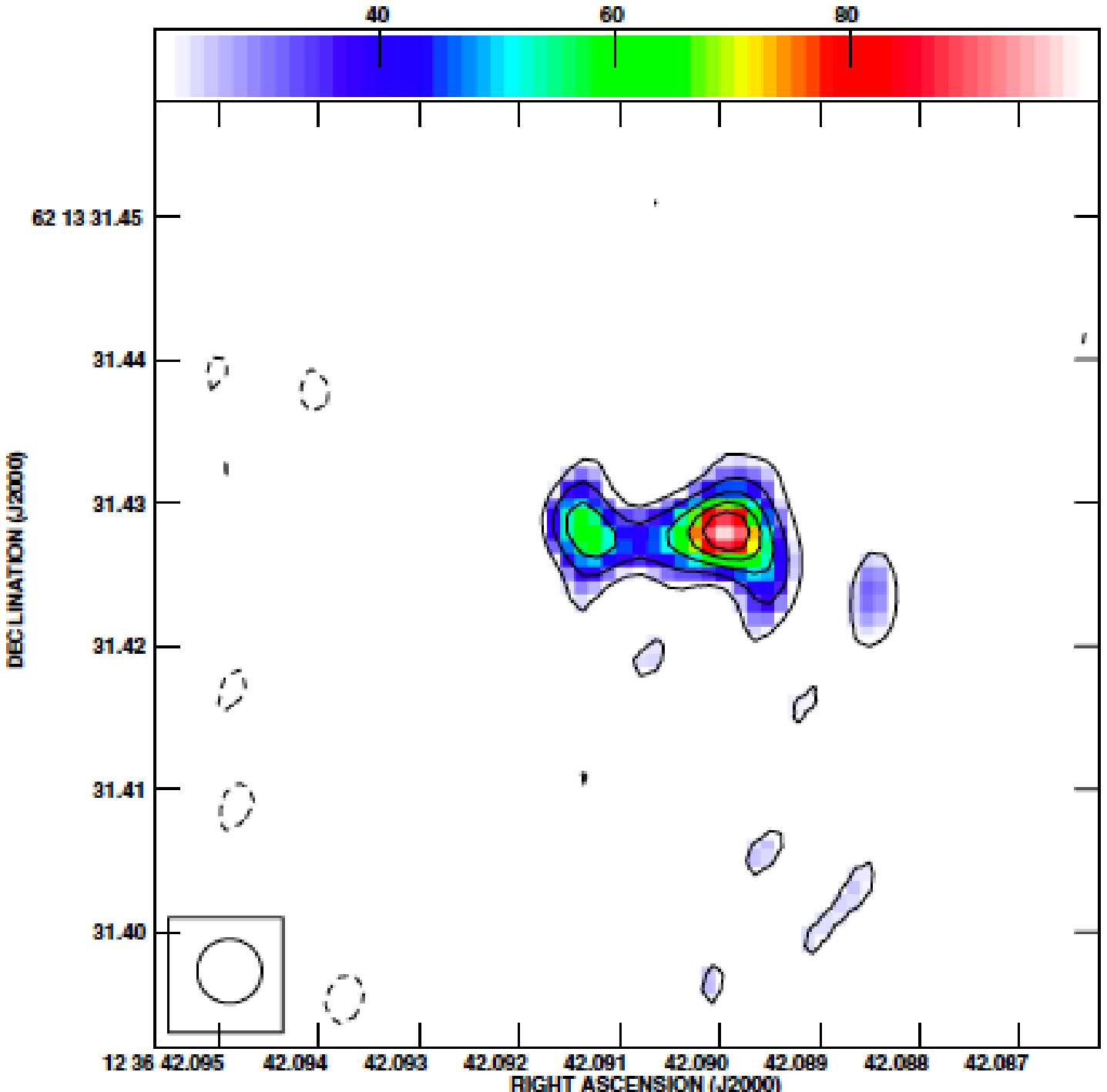}}
\hfill
\parbox[b]{55mm}{
\caption{{\it Left panel:} Radio emission on 200 mas scale at 1.4 GHz (colors, VLA and MERLIN) and 5 GHz (white contours, eMERLIN) for an ultraluminous galaxy at $z\sim 4.4$ in the GOODS-N region. Figure from \citep{Guidetti2013}. {\it Right panel:} VLBI image of 1.6 GHz radio emission with 4~mas resolution (both colors and contours) from the same galaxy confirming the presence of an AGN.  
Figure from \cite{Chi2013}.}
\label{fig:vlbi} }
\end{figure}

\subsection{Resolving AGN and SF with the SKA}

The power of the SKA  in unravelling the AGN/SF connections lies in its ability to trace the contribution from star-formation and AGN activity in individual galaxies out to high redshifts without any need to correct for poorly characterized selection effects induced by dust obscuration or orientation effects (e.g. Type 1 or Type 2 AGN). Morphological classification using high resolution radio observations is essential to determine the relative fraction of the emission generated by star-formation activity and nuclear activity. At faint, $\mu$Jy levels the majority of radio AGN are unresolved thus reliably separating the SF and AGN contributions  depends on our ability to distinguish compact AGN cores and inner jets at $\lesssim$kpc scales from star-forming disks and bulges with  typical sizes ranging from  1--10's~kpc. Figures~\ref{fig:merlin} and~\ref{fig:vlbi} are examples which illustrate the exceptional potential of high resolution (mas) radio observations as a discriminant between AGN and star-formation activity. In figure~\ref{fig:merlin} two intermediate/high redshift galaxies in the HDF North have been observed at 1.4 GHz from arcsec to mas scales. At VLBI scale an AGN core is detected in the first galaxy, while no VLBI detection is found for the second, supporting a merger scenario, over a core and jet one. This implies that in this object star formation processes are at work. Figure~\ref{fig:vlbi} shows a distant ($z\sim 4.4$) ultra-luminous infrared galaxy  identified in the GOODS-N region, and  interpreted as a dusty star-forming galaxy with an embedded weak AGN \citep[][]{Waddington1999}. eMERLIN and global VLBI observations ($>$ 1000~km baselines) were ultimately able to confirm the presence of an embedded AGN by resolving a jet and core structure separated by $\sim$70~pc. This may be an example of a high redshift ultra luminous infrared galaxy in which the high star-formation rate and the efficiency are enhanced by AGN jet activity \citep[e.g.][]{Silk2005}. This radio morphological identification is particularly  valuable in cases of highly obscured AGN activity which goes largely undetected at other wavelengths. High resolution radio observations have revealed highly compact AGN cores in submillimeter galaxies whose emission at near-infrared and X-ray wavelengths were completely devoid of AGN signatures \citep{Casey2009}. 

While high resolution observations at mid frequency (e.g. 1 GHz) will detect the extended steep spectrum emission from star-forming regions in an unbiased manner, these frequencies are biased against detecting the contribution from AGN with inverted or Gigaherz Peaked Spectra (GPS) whose spectra turn-over at lower frequencies due to either synchrotron self absorption or free-free absorption from an inhomogenous screen \citep{Odea,Marr,Orientigps}. High resolution multifrequency observations are thus vitally important for unambiguously separating star-formation and AGN emission in the presence of absorption effects.

%

%

\section{SKA surveys}

The SKA continuum science reference surveys are outlined in Prandoni and Seymour (this volume), which have been designed to meet a wide variety of science goals. Of the four proposed reference surveys those which are most relevant to the goal of understanding the star-formation AGN interplay is the three tier survey in band 1/2 at $\sim$1~GHz at high resolution with SKA1-MID and the two tier survey at band 5 at 10~GHz. The three tiers at $\sim$1~GHz range in depth  (1$\sigma$) from 1, 0.2 and 0.05~$\mu$Jy over areas of 1000--5000, 10--30 and 1~square degrees with a resolution of 0.5\arcsec using SKA1-MID. The luminosity functions for star-forming galaxies and AGN for these surveys are presented in Jarvis et al. (this volume) and  Smol{\v c}i{\'c} et al. (this volume). The luminosity functions demonstrate that these three tiers will probe star-formation rates of 0.5~M$_\odot$~yr$^{-1}$ at z$\sim$0.5 and 10~M$_\odot$~yr$^{-1}$ at redshifts z$\sim$1--2 and z$\sim$3--4. They are also shown to be sensitive to both radio loud and radio quiet AGN with luminosities of 10$^{22}$~W~Hz$^{-1}$ at z$\sim$0.5, 1--2 and 3--4 in tiers 1, 2 and 3 respectively, thus probing the co-evolution of AGN and galaxies from the local universe to the early stages of galaxy evolution. Furthermore the area surveyed is sufficiently large to study the variation in the AGN star-formation connection as a function of all the variables of interest which include the black hole and stellar mass, AGN accretion rate, star-formation rate, accretion mode, merger status and environment. 

In the case of radio-quiet AGN the luminosity limits refer to the compact emission generated by nuclear activity. As discussed earlier, there is much debate as to the relative contributions of star-formation and AGN activity to the total radio flux in these objects, if the AGN core constitutes a very small fraction of the total emission this will have important implications for our ability to probe the star-formation AGN connection down to the limiting star-formation rates of the proposed surveys. However recent work by \citet{White2014} indicates that the AGN contributes a non-negligable radio flux in radio-quiet AGN.

With 200~km baselines the SKA1 should achieve a minimum angular resolution of between 0.3-0.5\arcsec\  at 1.4~GHz (SKA1 band 2), which should be sufficient to resolve $\sim$ 4~kpc structures at z$\gtrsim$1. High resolution observations have already been used, with some success, to identify  AGN and SF emission in the $\mu$Jy radio population in the Hubble Deep Field North \citep{Muxlow2005,Guidetti2013}. MERLIN observations with resolutions of 0.2-0.5\arcsec\  reveal that the majority of these sources have subgalactic sizes, typically $\sim$1-1.2\arcsec, and that very few, 2\%, have double lobes surrounding a compact core (2/92). These sizes have now been confirmed for larger samples with recent e-MERLIN observations (Wrigley et al., in prep), while stacking experiments suggest similar size distributions down to levels of a few $\mu$Jy \citep{Muxlow2007}. Unambiguous classification of the dominant emission mechanism (AGN/SF) at these resolutions still frequently relies on infrared or spectral index information, thus even in the era of the SKA multiwavelength data will be essential to studies of continuum radio sources. Multiwavelength SED modelling, spectroscopy and information from X-ray, optical and infrared surveys will all aid in decoupling the contribution to the bolometric output from AGN and star-formation. These complementary datasets will enable us to fully exploit radio observations from SKA1. The key surveys at other wavelengths that will complement the radio continuum surveys with the SKA are thoroughly explored in Prandoni and Seymour (this volume), Jarvis et al. (this volume) and  Ciliegi et al. (this volume).

The higher resolution and spectral index information provided by higher frequency observations by SKA1 in bands 3,4 and 5 would also be extremely valuable for identifying AGN activity either via the presence of flat spectrum compact core emission or compact emission whose brightness temperature exceeds that of starbursts. The two proposed tiers at higher frequencies are necessarily limited to small areas due to the smaller field of view and reduced sensitivity of the array in band 5. We are proposing a shallower, high resolution (0.05\arcsec) survey over 0.5 square degrees to 0.3~$\mu$Jy/beam, and a deeper, lower resolution (0.1\arcsec) survey to 0.03~$\mu$Jy~beam$^{-1}$ over 30 square arcminutes. The depths are chosen to match the two shallower 1~GHz surveys assuming sources with spectral indices $S_{\nu}\propto \nu^{-1}$. These high frequency observations provide a means to morphologically or spectrally identify AGN cores which may be obscured at other wavelengths and are difficult to distinguish from star-formation in lower resolution radio observations. The observations will also be able to resolve individual star formation and accretion-related emitting components to allow detailed investigations of their interactions. Higher frequency observations also probe the thermal (free-free) emission dominated region of the radio spectrum of high redshift galaxies (z$\geq$2). This is a more direct indicator of SF activity than the synchrotron emission used at lower frequencies and can thus provide essential complementary information about high redshift star-formation activity and its relationship to AGN activity. The value of high frequency observations  are discussed in more detail in Murphy et al. (this volume).  

Our science goals require sufficient resolution to separate the AGN cores from star-forming disks and this will require much higher resolution, of the order of $<$0.1\arcsec, than currently planned with SKA1 at mid frequencies ($\sim$1~GHz). Only by combining high sensitivity with high (mas) resolution is it possible to disentangle the AGN and SF contributions at high redshifts. Studies along these lines will be conducted in the near future over small deep fields by combining deep $\mu$Jy observations at sub-arcsec resolution carried out with the JVLA at 5 GHz (Guidetti et al. in prep) with e-MERLIN observations, allowing angular resolutions on spatial scales of 50-100 mas (eMERGE survey,  \citealp{Muxlow2008}). Sub-$\mu$Jy sensitivities will be routinely reached by SKA on similar spatial scales, as baselines up to $\sim 100$~km (SKA1) and $\sim 1000$~km (SKA) will be incrementally implemented, according to the current design. However, as shown in Figs.~\ref{fig:merlin} and \ref{fig:vlbi}, only VLBI-like observations allow us to securely pinpoint AGN cores in sources at cosmological redshifts which calls for $10\times$ longer baselines ($\gg 1000$ km) for the full SKA. In the interim, VLBI capability with SKA1 would greatly improve our ability to pinpoint embedded AGN emission in high resolution follow-up observations.

The increased resolution and sensitivity of the full SKA will vastly improve our abilities to morphologically separate the contributions from star-formation and accretion processes within individual galaxies. At 1.4~GHz a survey to 3~nJy (1$\sigma$) at $\sim$0.03\arcsec\  resolution could detect resolved star formation at 50~M$_{\odot}$~yr$^{-1}$ out to z$\sim$2. The increased sensitivity at higher frequencies will also make it feasible to obtain high resolution spectral index information over larger areas of the sky, which can further aid in the AGN/SF decomposition. A deep SKA survey at 10~GHz to 3~nJy (1$\sigma$) could detect resolved star formation taking place at z$\sim$2 at 50~M$_{\odot}$~yr$^{-1}$ at a resolution of 0.1\arcsec, or to 100~M$_{\odot}$~yr$^{-1}$ at 0.07\arcsec. At 3~nJy we could detect low luminosity AGN cores of 10$^{19}$~W~Hz$^{-1}$ at z$\sim$0.5, these luminosities are comparable to radio cores detected in faint Seyferts in the local universe \citep{Nagar,Giroletti2009}. 

As resolution is key to the science goals laid out in this chapter, the usefulness of a 50\% SKA1 depends strongly on its maximum available baseline. Earlier progress could thus be achieved by deploying some fraction of the longest baselines at an earlier stage of construction. Alternatively, if the 50\%~SKA1 has only maximum baselines of 50~km the resolution  at 1~GHz is $\sim$1.5\arcsec, which is less than the planned resolution of the deep tier of the VLASS survey of 0.65\arcsec\  at $\sim$3~GHz (Murphy et al, 2014). Greater scientific gains could be made by conducting a high frequency, band 5, counterpart to the VLASS deep tier which could provide resolved spectral index information at a matched, or slightly higher resolution. The planned VLASS depth is 1.5~$\mu$Jy over 10~square degrees, a comparable depth at 10~GHz equates to 0.6~$\mu$Jy~beam$^{-1}$, assuming a spectral index of S$_\nu \propto \nu^{-0.7}$. A 50\% sensitivity SKA1 could survey to this depth at 10~GHz over 1~square degree in $\sim$1000 hours. 

Due its unique combination of high resolution imaging and sensitivity to obscured star-formation and AGN activity the SKA will be the leading facility to clarify the interplay between AGN and star-formation over a large fraction of cosmic time and thoughout the epoch of peak star formation and AGN activity.


\begin{thebibliography}{96}
\expandafter\ifx\csname natexlab\endcsname\relax\def\natexlab#1{#1}\fi


\bibitem[Afonso et al.(2006)]{Afonso2006} Afonso, J., Mobasher, 
B., Koekemoer, A., Norris, R.~P., \& Cram, L.\ 2006, \aj, 131, 1216 

\bibitem[{{Allen} {et~al.}(2006){Allen}, {Dunn}, {Fabian}, {Taylor}, \&
  {Reynolds}}]{Allen2006}
{Allen}, S.~W., {Dunn}, R.~J.~H., {Fabian}, A.~C. et al. 2006, \mnras, 372, 21

\bibitem[{{Antonucci}(1993)}]{Antonucci}
{Antonucci}, R. 1993, \araa, 31, 473

\bibitem[Appleton et al.(2004)]{Appleton} Appleton, P.~N., 
Fadda, D.~T., Marleau, F.~R., et al.\ 2004, \apjs, 154, 147 


\bibitem[{{Barnes}(1992)}]{Barnes1992}
{Barnes}, J.~E. 1992, \apj, 393, 484

\bibitem[{{Barnes} \& {Hernquist}(1996)}]{Barnes1996}
{Barnes}, J.~E. \& {Hernquist}, L. 1996, \apj, 471, 115

\bibitem[{{Bell}(2003)}]{Bell2003}
{Bell}, E.~F. 2003, \apj, 586, 794

\bibitem[{{Best} \& {Heckman}(2012)}]{Best2012}
{Best}, P.~N. \& {Heckman}, T.~M. 2012, \mnras, 421, 1569

\bibitem[{{Best} {et~al.}(2006){Best}, {Kaiser}, {Heckman}, \&
  {Kauffmann}}]{Best2006}
{Best}, P.~N., {Kaiser}, C.~R., {Heckman}, T.~M. et al. 2006,
  \mnras, 368, L67

\bibitem[{{Best} {et~al.}(2005){Best}, {Kauffmann}, {Heckman}, {Brinchmann},
  {Charlot}, {Ivezi{\'c}}, \& {White}}]{Best2005}
{Best}, P.~N., {Kauffmann}, G., {Heckman}, T.~M. et al. 2005, \mnras, 362, 25

\bibitem[{{Best} {et~al.}(2007){Best}, {von der Linden}, {Kauffmann},
  {Heckman}, \& {Kaiser}}]{Best2007}
{Best}, P.~N., {von der Linden}, A., {Kauffmann}, G. et al. 2007, \mnras, 379, 894

\bibitem[\protect\citeauthoryear{Beswick et 
al.}{2008}]{Beswick2008} Beswick R.~J., Muxlow T.~W.~B., Thrall H., 
Richards A.~M.~S., Garrington S.~T., 2008, MNRAS, 385, 1143 

\bibitem[{{Beswick}{et~al.}(2014){Beswick}}]{Beswickvol}{Beswick}, R.~J. et al. 2015, ``SKA studies of nearby galaxies: star-formation, accretion processes and molecular gas across all environments'', in proceedings of ``Advancing Astrophysics with the Square Kilometre Array", PoS(AASKA14)070


\bibitem[{{Bicknell} {et~al.}(2000){Bicknell}, {Sutherland}, {van Breugel},
  {Dopita}, {Dey}, \& {Miley}}]{Bicknell2000}
{Bicknell}, G.~V., {Sutherland}, R.~S., {van Breugel}, W.~J.~M. et al. 2000, \apj, 540, 678

\bibitem[{{Biggs} {et~al.}(2010){Biggs}, {Younger}, \& {Ivison}}]{Biggs2010}
{Biggs}, A.~D., {Younger}, J.~D., \& {Ivison}, R.~J. 2010, \mnras, 408, 342

\bibitem[{{B{\^i}rzan} {et~al.}(2004){B{\^i}rzan}, {Rafferty}, {McNamara},
  {Wise}, \& {Nulsen}}]{Birzan}
{B{\^i}rzan}, L., {Rafferty}, D.~A., {McNamara}, B.~R. et al. 2004, \apj, 607, 800

\bibitem[{{Bonzini} {et~al.}(2012){Bonzini}, {Mainieri}, {Padovani},
  {Kellermann}, {Miller}, {Rosati}, {Tozzi}, {Vattakunnel}, {Balestra},
  {Brandt}, {Luo}, \& {Xue}}]{Bonzini2012}
{Bonzini}, M., {Mainieri}, V., {Padovani}, P. et al. 2012, \apjs, 203, 15

\bibitem[{{Bonzini} {et~al.}(2013){Bonzini}, {Padovani}, {Mainieri},
  {Kellermann}, {Miller}, {Rosati}, {Tozzi}, \& {Vattakunnel}}]{Bonzini2013}
{Bonzini}, M., {Padovani}, P., {Mainieri}, V. et al. 2013, \mnras, 436, 3759

\bibitem[{{Bourne} {et~al.}(2011){Bourne}, {Dunne}, {Ivison}, {Maddox},
  {Dickinson}, \& {Frayer}}]{Bourne}
{Bourne}, N., {Dunne}, L., {Ivison}, R.~J. et al. 2011, \mnras, 410, 1155

\bibitem[{{Bower} {et~al.}(2006){Bower}, {Benson}, {Malbon}, {Helly}, {Frenk},
  {Baugh}, {Cole}, \& {Lacey}}]{Bower}
{Bower}, R.~G., {Benson}, A.~J., {Malbon}, R. et al. 2006, \mnras, 370, 645

\bibitem[{{Cano-D{\'{\i}}az} {et~al.}(2012){Cano-D{\'{\i}}az}, {Maiolino},
  {Marconi}, {Netzer}, {Shemmer}, \& {Cresci}}]{Cano-Diaz}
{Cano-D{\'{\i}}az}, M., {Maiolino}, R., {Marconi}, A. et al. 2012, \aap, 537, L8

\bibitem[{{Casey} {et~al.}(2009){Casey}, {Chapman}, {Muxlow}, {Beswick},
  {Alexander}, \& {Conselice}}]{Casey2009}
{Casey}, C.~M., {Chapman}, S.~C., {Muxlow}, T.~W.~B. et al. 2009, \mnras, 395, 1249

\bibitem[{{Cavagnolo} {et~al.}(2010){Cavagnolo}, {McNamara}, {Nulsen},
  {Carilli}, {Jones}, \& {B{\^i}rzan}}]{Cav}
{Cavagnolo}, K.~W., {McNamara}, B.~R., {Nulsen}, P.~E.~J. et al. 2010, \apj, 720, 1066

\bibitem[{{Chi} {et~al.}(2013){Chi}, {Barthel}, \& {Garrett}}]{Chi2013}
{Chi}, S., {Barthel}, P.~D., \& {Garrett}, M.~A. 2013, \aap, 550, A68

\bibitem[{{Cicone} {et~al.}(2014){Cicone}, {Maiolino}, {Sturm},
  {Graci{\'a}-Carpio}, {Feruglio}, {Neri}, {Aalto}, {Davies}, {Fiore},
  {Fischer}, {Garc{\'{\i}}a-Burillo}, {Gonz{\'a}lez-Alfonso},
  {Hailey-Dunsheath}, {Piconcelli}, \& {Veilleux}}]{Cicone2014}
{Cicone}, C., {Maiolino}, R., {Sturm}, E. et al. 2014, \aap, 562, A21

\bibitem[{{Ciliegi}{et~al.}(2014){Ciliegi}}]{Ciliegivol}{Ciliegi}, P. et al. 2015, ``Synergistic science with {\em EUCLID} and SKA: the nature and history of star formation'', in proceedings of ``Advancing Astrophysics with the Square Kilometre Array", PoS(AASKA14)150
  
  
\bibitem[{{Cisternas} {et~al.}(2011){Cisternas}, {Jahnke}, {Inskip},
  {Kartaltepe}, {Koekemoer}, {Lisker}, {Robaina}, {Scodeggio}, {Sheth},
  {Trump}, {Andrae}, {Miyaji}, {Lusso}, {Brusa}, {Capak}, {Cappelluti},
  {Civano}, {Ilbert}, {Impey}, {Leauthaud}, {Lilly}, {Salvato}, {Scoville}, \&
  {Taniguchi}}]{Cis}
{Cisternas}, M., {Jahnke}, K., {Inskip}, K.~J. et al.
  2011, \apj, 726, 57

\bibitem[{{Condon}(1992)}]{condon1992}
{Condon}, J.~J. 1992, \araa, 30, 575

\bibitem[{{Croft} {et~al.}(2006){Croft}, {van Breugel}, {de Vries}, {Dopita},
  {Martin}, {Morganti}, {Neff}, {Oosterloo}, {Schiminovich}, {Stanford}, \&
  {van Gorkom}}]{Croft2006}
{Croft}, S., {van Breugel}, W., {de Vries}, W., {Dopita}, M. et al. 2006, \apj, 647, 1040

\bibitem[{{Croston} {et~al.}(2008){Croston}, {Hardcastle}, {Birkinshaw},
  {Worrall}, \& {Laing}}]{Croston}
{Croston}, J.~H., {Hardcastle}, M.~J., {Birkinshaw}, M. et al. 2008, \mnras, 386, 1709

\bibitem[{{Croton} {et~al.}(2006){Croton}, {Springel}, {White}, {De Lucia},
  {Frenk}, {Gao}, {Jenkins}, {Kauffmann}, {Navarro}, \& {Yoshida}}]{Croton}
{Croton}, D.~J., {Springel}, V., {White}, S.~D.~M. et al. 2006, \mnras, 365, 11

\bibitem[{{Daddi} {et~al.}(2007){Daddi}, {Alexander}, {Dickinson}, {Gilli},
  {Renzini}, {Elbaz}, {Cimatti}, {Chary}, {Frayer}, {Bauer}, {Brandt},
  {Giavalisco}, {Grogin}, {Huynh}, {Kurk}, {Mignoli}, {Morrison}, {Pope}, \&
  {Ravindranath}}]{Daddi2007}
{Daddi}, E., {Alexander}, D.~M., {Dickinson}, M. et al. 2007, \apj,
  670, 173
  
\bibitem[\protect\citeauthoryear{Dasyra 
\& Combes}{2012}]{Dasyra2012} Dasyra K.~M., Combes F., 2012, A\&A, 541, LL7 
    

\bibitem[\protect\citeauthoryear{Delvecchio et 
al.}{2014}]{Delvecchio} Delvecchio I., et al. 2014, MNRAS, 439, 
2736 
  
  
\bibitem[{{Dey} {et~al.}(1997){Dey}, {van Breugel}, {Vacca}, \&
  {Antonucci}}]{Dey1997}
{Dey}, A., {van Breugel}, W., {Vacca}, W.~D., \& {Antonucci}, R. 1997, \apj,
  490, 698

\bibitem[{{Di Matteo} {et~al.}(2005){Di Matteo}, {Springel}, \&
  {Hernquist}}]{Dimatteo}
{Di Matteo}, T., {Springel}, V., \& {Hernquist}, L. 2005, \nat, 433, 604

\bibitem[\protect\citeauthoryear{Donoso, Best, 
\& Kauffmann}{2009}]{Donoso} Donoso E., Best P.~N., Kauffmann G., 2009, MNRAS, 392, 617 


\bibitem[{{Dubois} {et~al.}(2013){Dubois}, {Gavazzi}, {Peirani}, \&
  {Silk}}]{Dubois}
{Dubois}, Y., {Gavazzi}, R., {Peirani}, S., \& {Silk}, J. 2013, \mnras, 433,
  3297

\bibitem[{{Emonts} {et~al.}(2014){Emonts}, {Norris}, {Feain}, {Mao}, {Ekers},
  {Miley}, {Seymour}, {R{\"o}ttgering}, {Villar-Mart{\'{\i}}n}, {Sadler},
  {Carilli}, {Mahony}, {de Breuck}, {Stroe}, {Pentericci}, {van Moorsel},
  {Drouart}, {Ivison}, {Greve}, {Humphrey}, {Wylezalek}, \&
  {Tadhunter}}]{Emonts2014}
{Emonts}, B.~H.~C., {Norris}, R.~P., {Feain}, I. et al. 2014, \mnras, 438, 2898

\bibitem[{{Fabian} {et~al.}(2003){Fabian}, {Sanders}, {Allen}, {Crawford},
  {Iwasawa}, {Johnstone}, {Schmidt}, \& {Taylor}}]{Fabian}
{Fabian}, A.~C., {Sanders}, J.~S., {Allen}, S.~W. et al. 2003,
  \mnras, 344, L43

\bibitem[{{Fragile} {et~al.}(2004){Fragile}, {Murray}, {Anninos}, \& {van
  Breugel}}]{Fragile2004}
{Fragile}, P.~C., {Murray}, S.~D., {Anninos}, P., \& {van Breugel}, W. 2004,
  \apj, 604, 74

\bibitem[\protect\citeauthoryear{Gabor 
\& Bournaud}{2013}]{Gabor2013} Gabor J.~M., Bournaud F., 2013, MNRAS, 434, 606 
  
  
\bibitem[{{Gabor} \& {Bournaud}(2014)}]{Gabor}
{Gabor}, J.~M. \& {Bournaud}, F. 2014, \mnras, 441, 1615

\bibitem[{{Gaibler} {et~al.}(2012){Gaibler}, {Khochfar}, {Krause}, \&
  {Silk}}]{Gaibler2012}
{Gaibler}, V., {Khochfar}, S., {Krause}, M., \& {Silk}, J. 2012, \mnras, 425,
  438

\bibitem[{{Garrett}(2002)}]{Garrett2002}
{Garrett}, M.~A. 2002, \aap, 384, L19

\bibitem[{{Gaspari} {et~al.}(2013){Gaspari}, {Ruszkowski}, \&
  {Oh}}]{Gaspari2013}
{Gaspari}, M., {Ruszkowski}, M., \& {Oh}, S.~P. 2013, \mnras, 432, 3401

\bibitem[{{Gebhardt} {et~al.}(2000){Gebhardt}, {Bender}, {Bower}, {Dressler},
  {Faber}, {Filippenko}, {Green}, {Grillmair}, {Ho}, {Kormendy}, {Lauer},
  {Magorrian}, {Pinkney}, {Richstone}, \& {Tremaine}}]{Geb}
{Gebhardt}, K., {Bender}, R., {Bower}, G. et al. 2000, \apjl, 539, L13

\bibitem[{{Giroletti} \& {Panessa}(2009)}]{Giroletti2009}
{Giroletti}, M. \& {Panessa}, F. 2009, \apjl, 706, L260

\bibitem[{{Granato} {et~al.}(2004){Granato}, {De Zotti}, {Silva}, {Bressan}, \&
  {Danese}}]{Granato}
{Granato}, G.~L., {De Zotti}, G., {Silva}, L., et al.
  2004, \apj, 600, 580

\bibitem[{{Guidetti} {et~al.}(2013){Guidetti}, {Bondi}, {Prandoni}, {Beswick},
  {Muxlow}, {Wrigley}, {Smail}, \& {McHardy}}]{Guidetti2013}
{Guidetti}, D., {Bondi}, M., {Prandoni}, I. et al. 2013, \mnras, 432,
  2798

\bibitem[{{Guidetti} {et~al.}(2011){Guidetti}, {Laing}, {Bridle}, {Parma}, \&
  {Gregorini}}]{Guidetti2011}
{Guidetti}, D., {Laing}, R.~A., {Bridle}, A.~H. et al. 2011, \mnras, 413, 2525

\bibitem[{{Guidetti} {et~al.}(2012){Guidetti}, {Laing}, {Croston}, {Bridle}, \&
  {Parma}}]{Guidetti2012}
{Guidetti}, D., {Laing}, R.~A., {Croston}, J.~H. et al. 2012, \mnras, 423, 1335

\bibitem[{{Hardcastle} {et~al.}(2007){Hardcastle}, {Evans}, \&
  {Croston}}]{Hardcastle2007}
{Hardcastle}, M.~J., {Evans}, D.~A., \& {Croston}, J.~H. 2007, \mnras, 376,
  1849

\bibitem[{{H{\"a}ring} \& {Rix}(2004)}]{Har}
{H{\"a}ring}, N. \& {Rix}, H.-W. 2004, \apjl, 604, L89

\bibitem[{{Heckman} \& {Best}(2014)}]{Heckman2014}
{Heckman}, T. \& {Best}, P. 2014, \araa, 52 

\bibitem[\protect\citeauthoryear{Holt et al.}{2011}]{Holt} 
Holt J., Tadhunter C.~N., Morganti R., Emonts B.~H.~C., 2011, MNRAS, 410, 
1527

\bibitem[\protect\citeauthoryear{Hopkins 
\& Elvis}{2010}]{Hopkins2010} Hopkins P.~F., Elvis M., 2010, MNRAS, 401, 7 


\bibitem[{{Inskip} {et~al.}(2008){Inskip}, {Villar-Mart{\'{\i}}n}, {Tadhunter},
  {Morganti}, {Holt}, \& {Dicken}}]{Inskip2008}
{Inskip}, K.~J., {Villar-Mart{\'{\i}}n}, M., {Tadhunter}, C.~N. et al. 2008, \mnras, 386, 1797

\bibitem[\protect\citeauthoryear{Janssen et 
al.}{2012}]{Janssen} Janssen R.~M.~J., R{\"o}ttgering H.~J.~A., Best P.~N., Brinchmann J., 2012, A\&A, 541, AA62 
    
  
\bibitem[{{Jarvis} \& {Rawlings}(2004)}]{Jarvis2004}
{Jarvis}, M.~J. \& {Rawlings}, S. 2004, \nar, 48, 1173

\bibitem[{{Jarvis}{et~al.}(2014){Jarvis}}]{Jarvisvol}{Jarvis}, M.~J. et al. 2015, ``The star-formation history of the Universe with the SKA'', in proceedings of ``Advancing Astrophysics with the Square Kilometre Array", PoS(AASKA14)068


\bibitem[{{Kauffmann} {et~al.}(2008){Kauffmann}, {Heckman}, \&
  {Best}}]{Kauffmann2008}
{Kauffmann}, G., {Heckman}, T.~M., \& {Best}, P.~N. 2008, \mnras, 384, 953

\bibitem[{{Kim} {et~al.}(2012){Kim}, {Seo}, {Stone}, {Yoon}, \&
  {Teuben}}]{Kim2012}
{Kim}, W.-T., {Seo}, W.-Y., {Stone}, J.~M. et al.
  2012, \apj, 747, 60

\bibitem[{{Kimball} {et~al.}(2011){Kimball}, {Kellermann}, {Condon},
  {Ivezi{\'c}}, \& {Perley}}]{Kimball2011}
{Kimball}, A.~E., {Kellermann}, K.~I., {Condon}, J.~J. et al. 2011, \apjl, 739, L29

\bibitem[{{Klamer} {et~al.}(2004){Klamer}, {Ekers}, {Sadler}, \&
  {Hunstead}}]{Klamer2004}
{Klamer}, I.~J., {Ekers}, R.~D., {Sadler}, E.~M., \& {Hunstead}, R.~W. 2004,
  \apjl, 612, L97

\bibitem[{{Kocevski} {et~al.}(2012){Kocevski}, {Faber}, {Mozena}, {Koekemoer},
  {Nandra}, {Rangel}, {Laird}, {Brusa}, {Wuyts}, {Trump}, {Koo}, {Somerville},
  {Bell}, {Lotz}, {Alexander}, {Bournaud}, {Conselice}, {Dahlen}, {Dekel},
  {Donley}, {Dunlop}, {Finoguenov}, {Georgakakis}, {Giavalisco}, {Guo},
  {Grogin}, {Hathi}, {Juneau}, {Kartaltepe}, {Lucas}, {McGrath}, {McIntosh},
  {Mobasher}, {Robaina}, {Rosario}, {Straughn}, {van der Wel}, \&
  {Villforth}}]{Koc}
{Kocevski}, D.~D., {Faber}, S.~M., {Mozena}, M. et al. 2012, \apj, 744, 148

\bibitem[{{Kormendy} \& {Kennicutt}(2004)}]{Kormendy2004}
{Kormendy}, J. \& {Kennicutt}, Jr., R.~C. 2004, ARA\&A, 42, 603

\bibitem[{{Lacki} \& {Thompson}(2010)}]{Lackib}
{Lacki}, B.~C. \& {Thompson}, T.~A. 2010, \apj, 717, 196

\bibitem[{{Lacki} {et~al.}(2010){Lacki}, {Thompson}, \& {Quataert}}]{Lackia}
{Lacki}, B.~C., {Thompson}, T.~A., \& {Quataert}, E. 2010, \apj, 717, 1

\bibitem[{{Li} {et~al.}(2008{\natexlab{a}}){Li}, {Kauffmann}, {Heckman},
  {Jing}, \& {White}}]{Lib}
{Li}, C., {Kauffmann}, G., {Heckman}, T.~M., {Jing}, Y.~P., \& {White},
  S.~D.~M. 2008{\natexlab{a}}, \mnras, 385, 1903

\bibitem[{{Li} {et~al.}(2008{\natexlab{b}}){Li}, {Kauffmann}, {Heckman},
  {White}, \& {Jing}}]{Lia}
{Li}, C., {Kauffmann}, G., {Heckman}, T.~M. et al. 2008{\natexlab{b}}, \mnras, 385, 1915

\bibitem[{{Liu} {et~al.}(2013){Liu}, {Zakamska}, {Greene}, {Nesvadba}, \&
  {Liu}}]{Liu}
{Liu}, G., {Zakamska}, N.~L., {Greene}, J.~E. et al. 2013, \mnras, 436, 2576

\bibitem[{{Magorrian} {et~al.}(1998){Magorrian}, {Tremaine}, {Richstone},
  {Bender}, {Bower}, {Dressler}, {Faber}, {Gebhardt}, {Green}, {Grillmair},
  {Kormendy}, \& {Lauer}}]{Mag}
{Magorrian}, J., {Tremaine}, S., {Richstone}, D. et al. 1998, \aj, 115, 2285

\bibitem[{{Maiolino} {et~al.}(2012){Maiolino}, {Gallerani}, {Neri}, {Cicone},
  {Ferrara}, {Genzel}, {Lutz}, {Sturm}, {Tacconi}, {Walter}, {Feruglio},
  {Fiore}, \& {Piconcelli}}]{Maiolino}
{Maiolino}, R., {Gallerani}, S., {Neri}, R. et al. 2012, \mnras, 425, L66

  
\bibitem[\protect\citeauthoryear{Mao et al.}{2011}]{Mao2011} 
Mao M.~Y., Huynh M.~T., Norris R.~P. et al. 2011, ApJ, 731, 79 

\bibitem[\protect\citeauthoryear{Marr et al.}{2014}]{Marr} 
Marr J.~M., Perry T.~M., Read J. et al. 2014, ApJ, 
780, 178 

  
\bibitem[{{McNamara} {et~al.}(2011){McNamara}, {Rohanizadegan}, \&
  {Nulsen}}]{McNamara2011}
{McNamara}, B.~R., {Rohanizadegan}, M., \& {Nulsen}, P.~E.~J. 2011, \apj, 727,
  39

\bibitem[{{McNamara} {et~al.}(2000){McNamara}, {Wise}, {Nulsen}, {David},
  {Sarazin}, {Bautz}, {Markevitch}, {Vikhlinin}, {Forman}, {Jones}, \&
  {Harris}}]{McNamara}
{McNamara}, B.~R., {Wise}, M., {Nulsen}, P.~E.~J. et al. 2000, \apjl, 534, L135

\bibitem[\protect\citeauthoryear{Morganti et 
al.}{2013}]{Morganti2013} Morganti R., Fogasy J., Paragi Z., 
Oosterloo T., Orienti M., 2013, Sci, 341, 1082 


\bibitem[{{Mould} {et~al.}(2000){Mould}, {Ridgewell}, {Gallagher}, {Bessell},
  {Keller}, {Calzetti}, {Clarke}, {Trauger}, {Grillmair}, {Ballester},
  {Burrows}, {Krist}, {Crisp}, {Evans}, {Griffiths}, {Hester}, {Hoessel},
  {Holtzman}, {Scowen}, {Stapelfeldt}, {Sahai}, {Watson}, \&
  {Meadows}}]{Mould2000}
{Mould}, J.~R., {Ridgewell}, A., {Gallagher}, III, J.~S. et al. 2000, \apj, 536, 266

\bibitem[{{Murphy}(2009)}]{Murphyb}
{Murphy}, E.~J. 2009, \apj, 706, 482

\bibitem[{{Murphy} {et~al.}(2009){Murphy}, {Chary}, {Alexander}, {Dickinson},
  {Magnelli}, {Morrison}, {Pope}, \& {Teplitz}}]{Murphya}
{Murphy}, E.~J., {Chary}, R.-R., {Alexander}, D.~M. et al. 2009, \apj,
  698, 1380

\bibitem[{{Murphy}{et~al.}(2014){Murphy}}]{Murphyvol}{Murphy}, M. et al. 2015, ``The Astrophysics of Star Formation Across Cosmic Time at $\geq$10~GHz with the SKA'', in proceedings of ``Advancing Astrophysics with the Square Kilometre Array", PoS(AASKA14)085


\bibitem[{{Muxlow} {et~al.}(2005){Muxlow}, {Richards}, {Garrington},
  {Wilkinson}, {Anderson}, {Richards}, {Axon}, {Fomalont}, {Kellermann},
  {Partridge}, \& {Windhorst}}]{Muxlow2005}
{Muxlow}, T.~W.~B., {Richards}, A.~M.~S., {Garrington}, S.~T. et al. 2005, \mnras,
  358, 1159

\bibitem[\protect\citeauthoryear{Muxlow et al.}{2007}]{Muxlow2007} 
Muxlow T.~W.~B., Beswick R.~J., Thrall H. et al. 2007, ASPC, 380, 199 
  
\bibitem[{{Muxlow} {et~al.}(2008){Muxlow}, {Smail}, {McHardy}, \& {E-Merge
  Consortium}}]{Muxlow2008}
{Muxlow}, T., {Smail}, I., {McHardy}, I. et al. \& {E-Merge Consortium}. 2008, in The
  role of VLBI in the Golden Age for Radio Astronomy

\bibitem[\protect\citeauthoryear{Nagar et 
al.}{2002}]{Nagar} Nagar N.~M., Falcke H., Wilson A.~S., Ulvestad J.~S., 2002, A\&A, 392, 53 
  
  
\bibitem[\protect\citeauthoryear{O'Dea}{1998}]{Odea} O'Dea 
C.~P., 1998, PASP, 110, 493 


  
\bibitem[{{Orienti}{et~al.}(2014){Orienti}}]{Orientivol}{Orienti}, M. et al. 2015, ``The physics of the radio emission in the quiet side of the AGN population'', in proceedings of ``Advancing Astrophysics with the Square Kilometre Array", PoS(AASKA14)087

\bibitem[\protect\citeauthoryear{Orienti 
\& Dallacasa}{2014}]{Orientigps} Orienti M., Dallacasa D., 2014, MNRAS, 438, 463 



\bibitem[{{Padovani} {et~al.}(2009){Padovani}, {Mainieri}, {Tozzi},
  {Kellermann}, {Fomalont}, {Miller}, {Rosati}, \& {Shaver}}]{Padovani2009}
{Padovani}, P., {Mainieri}, V., {Tozzi}, P. et al. 2009, \apj, 694, 235

\bibitem[{{Padovani} {et~al.}(2011){Padovani}, {Miller}, {Kellermann},
  {Mainieri}, {Rosati}, \& {Tozzi}}]{Padovani2011}
{Padovani}, P., {Miller}, N., {Kellermann}, K.~I. et al. 2011, \apj, 740, 20
  
\bibitem[\protect\citeauthoryear{Pedlar et al.}{1990}]{Pedlar} 
Pedlar A., Ghataure H.~S., Davies R.~D. et al. 1990, MNRAS, 246, 477 
  

\bibitem[{{Pizzolato} \& {Soker}(2010)}]{Pizzolato2010}
{Pizzolato}, F. \& {Soker}, N. 2010, \mnras, 408, 961

\bibitem[{{Pounds} \& {King}(2013)}]{Pounds}
{Pounds}, K.~A. \& {King}, A.~R. 2013, \mnras, 433, 1369

\bibitem[Prandoni et 
al.(2001)]{Prandoni2001} Prandoni, I., Gregorini, L., Parma, P., et al.\ 2001, \aap, 369, 787 


\bibitem[{{Prandoni} {et~al.}(2010){Prandoni}, {de Ruiter}, {Ricci}, {Parma},
  {Gregorini}, \& {Ekers}}]{Prandoni2010}
{Prandoni}, I., {de Ruiter}, H.~R., {Ricci}, R. et al. 2010, \aap, 510, A42

  
\bibitem[{{Prandoni}{and Seymour}(2014){Prandoni}}]{Prandonivol}{Prandoni} I. \& {Seymour} N. 2015, ``Revealing the Physics and Evolution of Galaxies and Galaxy clusters with SKA Continuum Surveys'', in proceedings of ``Advancing Astrophysics with the Square Kilometre Array", PoS(AASKA14)067

\bibitem[{{Puchwein} \& {Springel}(2013)}]{Puch}
{Puchwein}, E. \& {Springel}, V. 2013, \mnras, 428, 2966

\bibitem[{{Rafferty} {et~al.}(2006){Rafferty}, {McNamara}, {Nulsen}, \&
  {Wise}}]{Rafferty}
{Rafferty}, D.~A., {McNamara}, B.~R., {Nulsen}, P.~E.~J. et al.
  2006, \apj, 652, 216

\bibitem[{{Reichard} {et~al.}(2009){Reichard}, {Heckman}, {Rudnick},
  {Brinchmann}, {Kauffmann}, \& {Wild}}]{Reichard2009}
{Reichard}, T.~A., {Heckman}, T.~M., {Rudnick}, G. et al. 2009, \apj, 691, 1005

\bibitem[{{Sargent} {et~al.}(2010){Sargent}, {Schinnerer}, {Murphy}, {Carilli},
  {Helou}, {Aussel}, {Le Floc'h}, {Frayer}, {Ilbert}, {Oesch}, {Salvato},
  {Smol{\v c}i{\'c}}, {Kartaltepe}, \& {Sanders}}]{Sargent}
{Sargent}, M.~T., {Schinnerer}, E., {Murphy}, E. et al. 2010, \apjl, 714, L190

\bibitem[{{Schawinski} {et~al.}(2012){Schawinski}, {Simmons}, {Urry},
  {Treister}, \& {Glikman}}]{Sch}
{Schawinski}, K., {Simmons}, B.~D., {Urry}, C.~M. et al. \mnras, 425, L61

\bibitem[{{Scott} {et~al.}(2013){Scott}, {Graham}, \& {Schombert}}]{Scott}
{Scott}, N., {Graham}, A.~W., \& {Schombert}, J. 2013, \apj, 768, 76

\bibitem[Seymour et al.(2008)]{Seymour2008} Seymour, N., Dwelly, 
T., Moss, D., et al.\ 2008, \mnras, 386, 1695 


\bibitem[{{Silk}(2005)}]{Silk2005}
{Silk}, J. 2005, \mnras, 364, 1337

\bibitem[{{Simpson} {et~al.}(2006){Simpson}, {Mart{\'{\i}}nez-Sansigre},
  {Rawlings}, {Ivison}, {Akiyama}, {Sekiguchi}, {Takata}, {Ueda}, \&
  {Watson}}]{Simpson2006}
{Simpson}, C., {Mart{\'{\i}}nez-Sansigre}, A., {Rawlings}, S., et al.
  2006, \mnras, 372, 741

\bibitem[{{Simpson} {et~al.}(2013){Simpson}, {Westoby}, {Arumugam}, {Ivison},
  {Hartley}, \& {Almaini}}]{Simpson2013}
{Simpson}, C., {Westoby}, P., {Arumugam}, V. et al. 2013, \mnras, 433, 2647

\bibitem[{{Smol{\v c}i{\'c}} {et~al.}(2008){Smol{\v c}i{\'c}}, {Schinnerer},
  {Scodeggio}, {Franzetti}, {Aussel}, {Bondi}, {Brusa}, {Carilli}, {Capak},
  {Charlot}, {Ciliegi}, {Ilbert}, {Ivezi{\'c}}, {Jahnke}, {McCracken},
  {Obri{\'c}}, {Salvato}, {Sanders}, {Scoville}, {Trump}, {Tremonti}, {Tasca},
  {Walcher}, \& {Zamorani}}]{Smolcic2008}
{Smol{\v c}i{\'c}}, V., {Schinnerer}, E., {Scodeggio}, M. et al. 2008, \apjs, 177, 14

\bibitem[{{Smol{\v c}i{\'c}}{et~al.}(2014){Smol{\v c}i{\'c}}}]{smolvol}{Smol{\v c}i{\'c}}, V. et al. 2015, ``Exploring AGN Activity over Cosmic Time'', in proceedings of ``Advancing Astrophysics with the Square Kilometre Array", PoS(AASKA14)069
  
  
\bibitem[{{Somerville} {et~al.}(2008){Somerville}, {Hopkins}, {Cox},
  {Robertson}, \& {Hernquist}}]{Somerville}
{Somerville}, R.~S., {Hopkins}, P.~F., {Cox}, T.~J., et al. 2008, \mnras, 391, 481

\bibitem[{{Springel} {et~al.}(2005){Springel}, {Di Matteo}, \&
  {Hernquist}}]{Springel}
{Springel}, V., {Di Matteo}, T., \& {Hernquist}, L. 2005, \apjl, 620, L79

\bibitem[{{Steinbring}(2014)}]{Steinbring2014}
{Steinbring}, E. 2014, \aj, 148, 10

\bibitem[{{Sturm} {et~al.}(2011){Sturm}, {Gonz{\'a}lez-Alfonso}, {Veilleux},
  {Fischer}, {Graci{\'a}-Carpio}, {Hailey-Dunsheath}, {Contursi}, {Poglitsch},
  {Sternberg}, {Davies}, {Genzel}, {Lutz}, {Tacconi}, {Verma}, {Maiolino}, \&
  {de Jong}}]{Sturm}
{Sturm}, E., {Gonz{\'a}lez-Alfonso}, E., {Veilleux}, S. et al. 2011, \apjl, 733, L16

\bibitem[{{Thomson} {et~al.}(2014){Thomson}, {Ivison}, {Simpson}, {Swinbank},
  {Smail}, {Arumugam}, {Alexander}, {Beelen}, {Brandt}, {Chandra},
  {Dannerbauer}, {Greve}, {Hodge}, {Ibar}, {Karim}, {Murphy}, {Schinnerer},
  {Sirothia}, {Walter}, {Wardlow}, \& {van der Werf}}]{Thomson}
{Thomson}, A.~P., {Ivison}, R.~J., {Simpson}, J.~M. et al. 2014, ArXiv:1404.7128T

\bibitem[{{Treister} {et~al.}(2012){Treister}, {Schawinski}, {Urry}, \&
  {Simmons}}]{Treister2012}
{Treister}, E., {Schawinski}, K., {Urry}, C.~M. et al. 2012,
  \apjl, 758, L39

\bibitem[{{Tremaine} {et~al.}(2002){Tremaine}, {Gebhardt}, {Bender}, {Bower},
  {Dressler}, {Faber}, {Filippenko}, {Green}, {Grillmair}, {Ho}, {Kormendy},
  {Lauer}, {Magorrian}, {Pinkney}, \& {Richstone}}]{Tremaine}
{Tremaine}, S., {Gebhardt}, K., {Bender}, R. et al. 2002, \apj, 574, 740

\bibitem[{{Urrutia} {et~al.}(2008){Urrutia}, {Lacy}, \& {Becker}}]{Urrutia2008}
{Urrutia}, T., {Lacy}, M., \& {Becker}, R.~H. 2008, \apj, 674, 80

\bibitem[{{Voit} \& {Donahue}(2005)}]{Voit2005}
{Voit}, G.~M. \& {Donahue}, M. 2005, \apj, 634, 955

\bibitem[{{Waddington} {et~al.}(1999){Waddington}, {Windhorst}, {Cohen},
  {Partridge}, {Spinrad}, \& {Stern}}]{Waddington1999}
{Waddington}, I., {Windhorst}, R.~A., {Cohen}, S.~H. et al. 1999, \apjl, 526, L77

\bibitem[{{White} \& {Frenk}(1991)}]{White1991}
{White}, S.~D.~M. \& {Frenk}, C.~S. 1991, \apj, 379, 52

\bibitem[\protect\citeauthoryear{White et al.}{2014}]{White2014} 
White S.~V., Jarvis M.~J., H{\"a}u{\ss}ler B. et al. 2014, arXiv, 
arXiv:1410.3892 

\bibitem[{{Wild} {et~al.}(2007){Wild}, {Kauffmann}, {Heckman}, {Charlot},
  {Lemson}, {Brinchmann}, {Reichard}, \& {Pasquali}}]{Wild2007}
{Wild}, V., {Kauffmann}, G., {Heckman}, T. et al. 2007, \mnras, 381, 543

\bibitem[{{Yun} {et~al.}(2001){Yun}, {Reddy}, \& {Condon}}]{Yun}
{Yun}, M.~S., {Reddy}, N.~A., \& {Condon}, J.~J. 2001, \apj, 554, 803

\end{thebibliography}

\end{document}